\documentclass[twocolumn,aps,prl,showpacs]{revtex4}
\usepackage{amsmath}
\usepackage{graphicx}
\usepackage{dcolumn}
\usepackage{bm}
\usepackage{float}

\begin{document}

\draft
\emph{}
\title{Non-equilibrium Kinetics of the Transformation of Liquids into Physical Gels}
\author{Jos\'e Manuel Olais-Govea, Leticia L\'opez-Flores*, Mart\'in Ch\'avez-P\'aez, and
Magdaleno Medina-Noyola}

\address{Instituto de F\'{\i}sica {\sl ``Manuel Sandoval Vallarta"},
Universidad Aut\'{o}noma de San Luis Potos\'{\i}, \'{A}lvaro
Obreg\'{o}n 64, 78000 San Luis Potos\'{\i}, SLP, M\'{e}xico}

\date{\today}

\begin{abstract}
A major stumbling block for statistical physics and materials science has been the lack of a universal principle that allows us to understand and predict elementary structural, morphological, and dynamical properties of non-equilibrium amorphous states of matter. The recently-developed non-equilibrium self-consistent generalized Langevin equation (NE-SCGLE) theory, however, has been shown to provide a fundamental tool for the understanding of  the most essential features of the transformation of liquids into amorphous solids, such as their aging kinetics or their dependence on the protocol of fabrication. In this work we focus on the predicted kinetics of one of the main fingerprints of the formation of gels by arrested spinodal decomposition of suddenly and deeply quenched simple liquids, namely, the arrest of structural parameters associated with the morphological evolution from the initially uniform fluid, to the dynamically arrested sponge-like amorphous material. The comparison of the theoretical predictions (based on a simple specific model system), with simulation and experimental data measured on similar but more complex materials, suggests the universality of the predicted scenario.

\end{abstract}


\maketitle

In spite of its relevance, there seems to be no universal principle that explains how Boltzmann's postulate  $S=k_B \ln W$ operates for non-equilibrium conditions, such that it predicts, for example, the transformation of liquids into non-equilibrium amorphous solids such as glasses, gels, etc. \cite{angellreview1,zaccarellireviewgels}, in terms of molecular interactions. For instance,  quenching a simple liquid to inside its gas-liquid spinodal region, normally leads
to the full phase separation \cite{cahnhilliard,cook,furukawa,langer,dhont}. Under some
conditions, however, this process may be interrupted when the
denser phase solidifies as an amorphous sponge-like
non-equilibrium bicontinuous structure with statistically well-defined spatial
heterogeneities, whose final mean size $\xi_a$ depends on the density and final temperature of the quench \cite{lu,sanz,gibaud,foffi,helgeson,davela,royall,heyeslodge,testardjcp}.

This process, referred to as \emph{arrested} spinodal decomposition, is revealed by the  development of a peak at small wave-vectors in the non-equilibrium structure factor $S(k;t)\equiv \langle\delta n (\textbf{k},t) \delta n (-\textbf{k},t)\rangle$ of many real
\cite{lu,sanz,gibaud,foffi,helgeson,davela,royall} and simulated \cite{heyeslodge,testardjcp,sanz,foffi} gel-forming liquids. Its most remarkable kinetic fingerprint is the fact that the position $k_{max}(t)$ of this non-equilibrium peak decreases with waiting time $t$ until the mean size $\xi (t)=2\pi/k_{max} (t)$ of these heterogeneities saturates at the finite ``arrested'' value  $\xi_a$.

Most of the previous experimental and simulation reports \cite{lu,sanz,gibaud,foffi,helgeson,heyeslodge,testardjcp}
acknowledge the notable absence of a fundamental predictive theory that explains the universal and the specific features of the evolution of non-equilibrium properties, such as $S(k;t)$. It is not clear, for instance \cite{berthierreview}, how to extend the classical theory of spinodal decomposition \cite{cahnhilliard, cook, furukawa, langer, dhont} to include the possibility of dynamic arrest, or how to incorporate the characteristic non-stationarity of spinodal decomposition in existing theories of glassy behavior \cite{goetzebook,langer1}. Thus, for example, in spite of its impressive predictive power, illustrated by the existence of attractive glasses in systems with short-ranged attractions \cite{Bergenholtz,foffi0},  mode-coupling theory remains in essence an equilibrium theory, unable to describe non-stationary processes such as aging.

The present work starts with the assumption that the manner how Boltzmann's postulate explains non-equilibrium states  is provided by Onsager's description of irreversible processes and thermal fluctuations, adequately extended to include spatial and temporal non-localities, as well as genuine non-equilibrium conditions  \cite{nescgle0,nescgle1}. Its application as a generic theory of irreversible processes in liquids, referred to as the non-equilibrium self-consistent generalized Langevin equation (NE-SCGLE) theory \cite{nescgle1,nescgle2,nescgle3}, seems to provide the long-awaited fundamental framework to understand the phenomenology of structural glasses and gels in terms of their specific molecular constitution. Judging by its predictions  \cite{nescgle4,nescgle5,nescgle6}, this non-equilibrium theory represents a major opportunity for progress in the science and engineering of these materials. The main purpose of this short communication is to report that this theory provides, in particular,  a vivid description of the  \emph{kinetics} of the \emph{structural} transformation of simple liquids into physical gels by arrested spinodal decomposition, \emph{a feature never before achieved by any other theory}. Here we also illustrate the universal nature of the main qualitative features of these predictions, by comparing them with observations in a wide variety of experimental realizations.

Let us start by considering a generic (``Lennard-Jones--like'') monocomponent simple liquid, with pairwise repulsive core, plus weaker longer-ranged attractive interactions. Assume that we subject this system to an \emph{instantaneous isochoric} cooling at waiting time $t=0$, from an initial temperature $T_i$ to a final temperature $T_f$. We then let the system relax under isochoric conditions and in the absence of applied external fields. As explained in Ref. \cite{nescgle6} and summarized here in the Supplemental Material (SM), the NE-SCGLE theory describes the spontaneous response of this system in terms of the non-equilibrium static structure factor $S(k;t)$, whose time-evolution equation  for $t>0$ reads
\begin{equation}
\frac{\partial S(k;t)}{\partial t} = -2k^2 D_0
b(t)n\mathcal{E}_f(k) \left[S(k;t)
-1/n\mathcal{E}_f(k)\right], \label{relsigmadif2pp}
\end{equation}
where  $D_0$ is the short-time self-diffusion coefficient \cite{atomic0,atomic1}. In this equation $\mathcal{E}_f(k)\equiv \mathcal{E}(k;\overline{n},T_f)$ is the Fourier transform of the  thermodynamic functional derivative
$\mathcal{E}[{\bf r},{\bf r}';n,T]\equiv \left[ {\delta \beta\mu
[{\bf r};n,T]}/{\delta n({\bf r}')}\right]$, evaluated at the
uniform density and temperature profiles $n(\textbf{r}) = n$ and
$T(\textbf{r}) = T$, in which case it can be written as
$\mathcal{E}[\textbf{r},\textbf{r}';n,T]=
\mathcal{E}(|\textbf{r}-\textbf{r}'|;n,T)=\delta
(\textbf{r}-\textbf{r}')/n-c(|\textbf{r}-\textbf{r}'|;n,T)$ or, in
Fourier space, as $\mathcal{E}(k;n, T)=1/n-c(k;n, T)$, where
$c(r;n,T)$ is the so-called direct correlation function.

The key ingredient in Eq. (\ref{relsigmadif2pp}) is the $t$-dependent mobility function $b(t)$, which is in reality a \emph{functional} of $S(k;t)$, and this introduces strong non-linearities in Eq. (\ref{relsigmadif2pp}). As a result, besides the stationary solutions $\lim_{t\to \infty} S(k;t)=S^{eq}(k;n,T_f) \equiv 1/n\mathcal{E}(k;n,T_f)$, representing ordinary thermodynamic equilibrium states, Eq. (\ref{relsigmadif2pp}) also predicts dynamically-arrested stationary solutions $S_a(k)$ when $\lim_{t\to \infty} b(t)=0$. These arrested solutions have in fact been calculated for soft-sphere liquids \cite{nescgle3,nescgle6}, and shown to represent  ``repulsive'' glasses.

\begin{figure}
\begin{center}
\includegraphics[scale=.25]{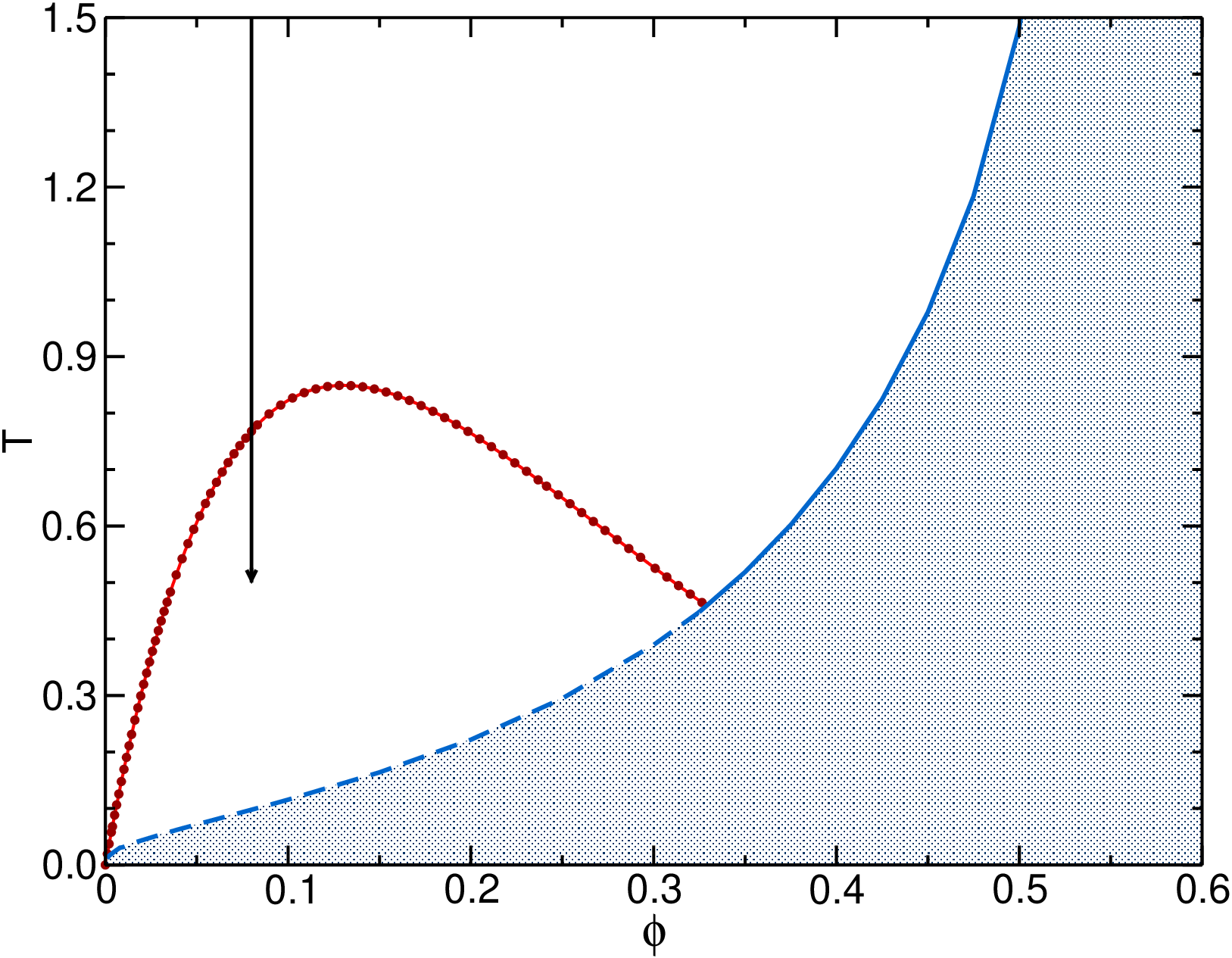}
\caption{\emph{Non-equilibrium} phase diagram of the HSAY ($z=2$) liquid \cite{nescgle5}.  At high temperatures and densities, cooling this liquid drives it through the transition (blue solid line) to non-equilibrium repulsive glasses \cite{nescgle3}. At intermediate and low densities and temperatures the NE-SCGLE theory predicts that (\emph{i}) the spinodal line $T_s(\phi)$ (red doted line) is a frontier between equilibrium and non-ergodic states, (\emph{ii}) the  liquid-glass transition line penetrates inside the spinodal region as a glass-glass transition line (dashed blue line), (\emph{iii}) below the composed (solid and dashed) blue line there exists a continuous region of porous repulsive glasses (shaded region). The vertical arrow represents the instantaneous temperature quench referred to in Fig. \ref{Fig1}, from an initial high temperature $T_i = 1.5$, to a final temperature $T_f=0.5$ along the isochore $\phi=0.08$.} \label{Fig0}
\end{center}
\end{figure}

A far more interesting prediction arises, however,  when Eq. (\ref{relsigmadif2pp}) is applied to the description of the spinodal decomposition of Lennard-Jones--like simple liquids \cite{nescgle5}, illustrated here for analytical simplicity with the hard-sphere plus attractive Yukawa (HSAY) model, defined by the pair potential
\begin{equation}
u(r)=
\begin{cases}
\infty, & r < \sigma; \\
-\epsilon \frac{\exp[-z(r/\sigma -1)]}{(r/\sigma )}, & r>\sigma.
\end{cases}
\label{yukawa}
\end{equation}
For given $\sigma$, $\epsilon$, and $z$, the state space of this system is spanned by the dimensionless number density $[n\sigma^3]$ and temperature $[k_BT/\epsilon]$, denoted simply as
$n$ and $T$ (i.e., we shall use $\sigma$ as the unit of length, and $\epsilon/k_B$ as the unit of temperature); we also refer to the hard-sphere volume fraction $\phi\equiv \pi n/6$. The dimensionless time $[D_0t/\sigma^2]$ will also be denoted simply as $t$. The main findings of Ref. \cite{nescgle5}, which only analyzed the asymptotic \emph{stationary} solutions of Eq. (\ref{relsigmadif2pp}), are summarized in Fig. \ref{Fig0}.

In contrast, the present work focuses on the detailed kinetics of $S(k;t)$ provided by the full solution of Eq. (\ref{relsigmadif2pp}) for the same model system subjected to instantaneous isochoric quenches illustrated by the vertical arrow in Fig. \ref{Fig0}. For this, we adopt a van der Waals (vdW) approximation for the Helmholtz free energy (see SM for details), which provides the approximate thermodynamic input $\mathcal{E}(k;n,T_f)$ needed in Eq. (1). This quench drives the system well inside the spinodal region, where no solution $S^{eq}(k;n,T_f)$ exists that corresponds to \emph{spatially uniform} equilibrium states.

\begin{figure}
\begin{center}
\includegraphics[scale=.25]{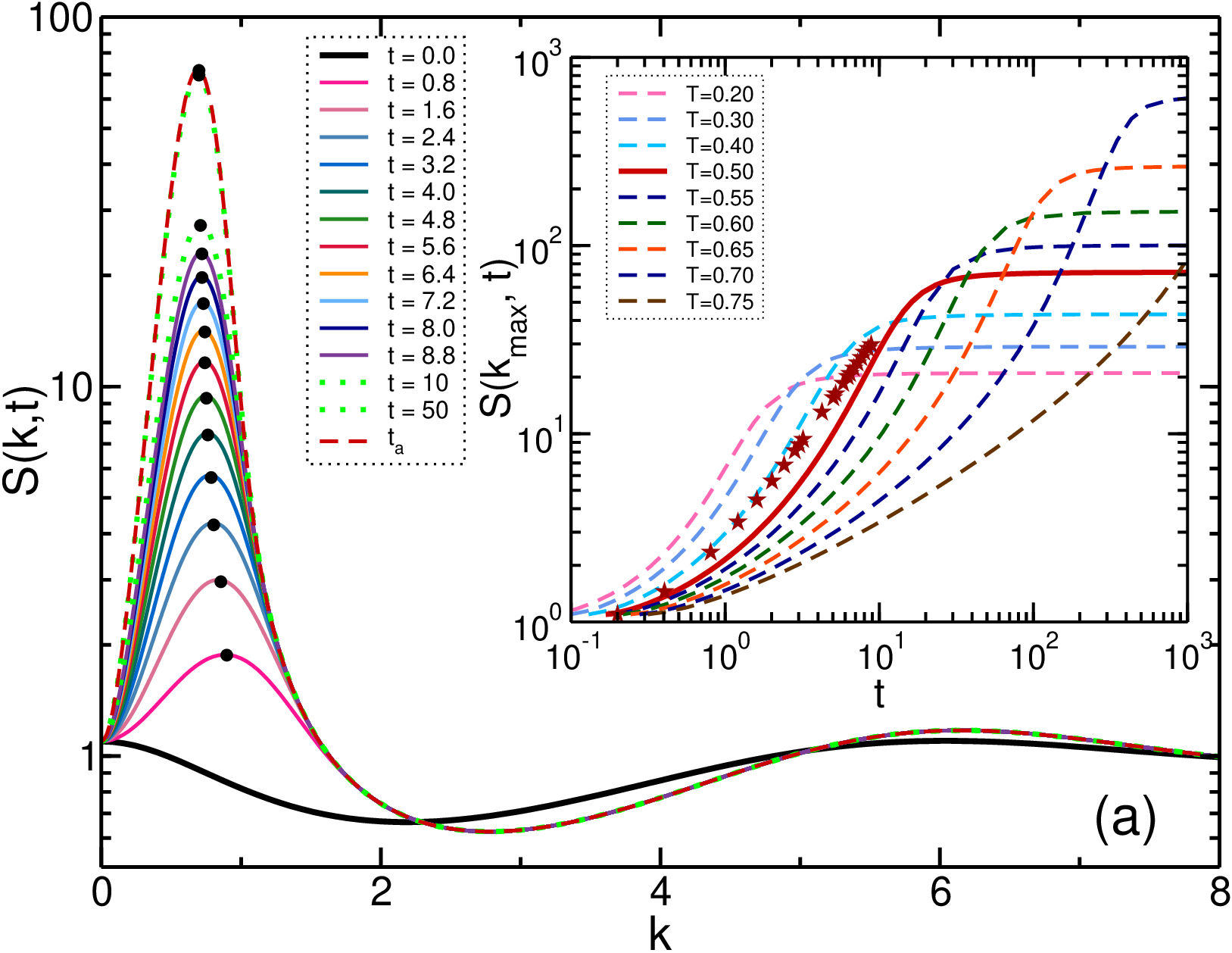}
\includegraphics[scale=.25]{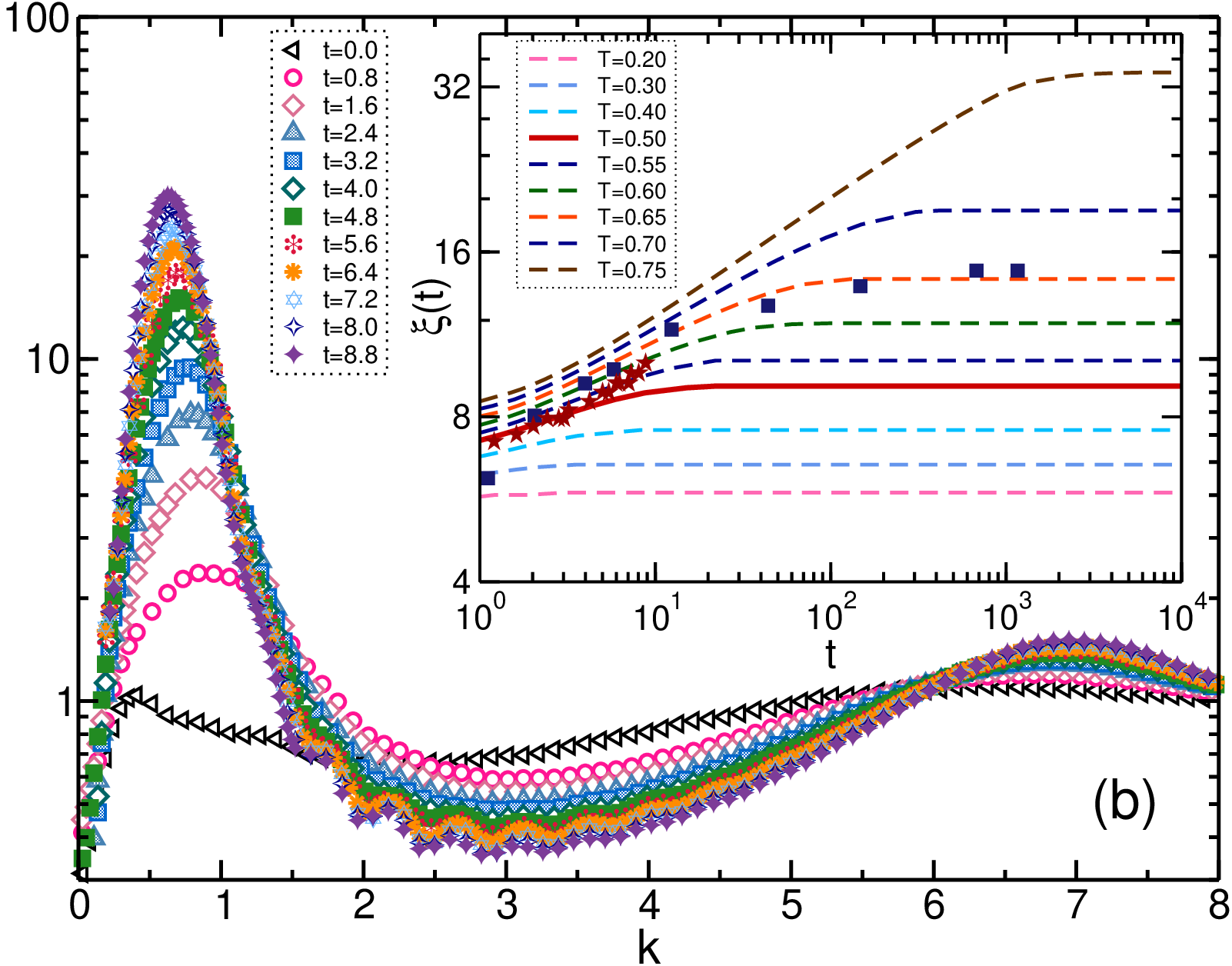}
\caption{Snapshots of the (a) theoretical prediction  and (b) Brownian-dynamics simulation, of the evolution of $S(k;t)$ from its initial state $S(k;t=0)$ towards its asymptotic arrested value $S_a(k)$ for (a) the quench indicated by the arrow in Fig. \ref{Fig0}, and (b) for the corresponding simulated quench (see text) at the same waiting times. The dark dots in (a) illustrate the evolution of the height $S_{max}(t)$ and position $k_{max}(t)$ of the small-$k$ peak of $S(k;t)$. The solid line of the insets shows the theoretical evolution (a) of $S_{max}(t)$ and (b) of the the wave-length $\xi (t)=2\pi/k_{max} (t)$ for this quench, whereas the other lines correspond to similar quenches differing only in the final temperature $T$ of the quench. The stars are the corresponding results of the simulation in the main panel of (b). The solid squares are experimental results measured in the gelling lysozyme solutions reported in Fig. 2(d) of Ref.\cite{gibaud}.
} \label{Fig1}
\end{center}
\end{figure}

Thus, the only possible uniform stationary solution is the non-equilibrium $S_a(k)$, and in Fig. \ref{Fig1}(a) we present the one corresponding to the quench represented by the arrow in Fig.  \ref{Fig0}, together with a sequence of snapshots describing the transient $S(k;t)$, which starts from the chosen initial value $S(k;t=0)=S_i(k)\equiv S^{eq}(k;\phi,T_i)$ and ends at $S_a(k)$. The first feature to notice in the structural kinetics illustrated by these snapshots is the fast but moderate (and  rather uneventful) growth of the main peak of $S(k;t)$ at $k\approx 2\pi$, compared with the dramatic development of the non-equilibrium spinodal decomposition peak at smaller wave-vectors, whose height $S_{max}(t)$ increases,  and whose position $k_{max} (t)$ decreases, until saturating at the finite values $S^a_{max}$ and  $k_a$, corresponding to $S_a(k)$ (red dashed line).

This kinetic process is summarized in the insets of  Figs. \ref{Fig1}(a) and \ref{Fig1}(b) by the solid lines, which illustrates the evolution of $S_{max}(t)$ and  of the wave-length $\xi (t)=2\pi/k_{max} (t)$ associated  with $k_{max}(t)$. The other (dashed) lines in both insets correspond to additional  processes that differ only in the depth of the quench, i.e., in the final temperature $T$. The comparison of these illustrative results indicates, for example, that deeper quenches lead to a faster increase (but also earlier arrest) of $S_{max}(t)$. The non-equilibrium evolution of  $\xi (t)$ bears an important morphological and kinetic significance, since it describes the growth and subsequent arrest of the mean size of the spinodal heterogeneities. As illustrated in this inset, $\xi (t)$ is predicted to increase with waiting time $t$ and to asymptotically saturate at the finite ``arrested'' value  $\xi_a=2\pi/k_a$. This maximum size $\xi_a(T)$ depends on the depth $T$ of the quench, and as discussed in Ref. \cite{nescgle5}, it is finite for $T$ smaller than the spinodal temperature $T_s$, but diverges when $T$ reaches $T_s$ from below. Thus, although for $T<T_s$ the emergence of dynamic arrest cancels the possibility of long-time asymptotic divergence of $\xi (t)$, before its saturation $\xi(t)$ appears to follow an apparent algebraic functional form  $\xi(t) \propto t^{\alpha}$ within a limited time-interval, with an exponent $\alpha$ that decreases with  the depth of the quench, attaining its maximum value when $T$ approaches $T_s$. We have verified that this predicted scenario is qualitatively independent of $\phi$, and even of the range $z^{-1}$ of the attractive term of the pair potential. Furthermore, we  also checked its independence on the specific form of the attractive potential, by repeating the same calculations for the hard sphere plus square well (HSSW) model liquid (in which the Yukawa tail is  substituted by a square well).

To determine to what extent a quantitative comparison can be established with the actual behaviour of the HSAY fluid, we performed non-equilibrium Brownian dynamics (BD) simulations on this precise model system to simulate a quench along the isochore $\phi=0.1$ from $T_i=2.0$ to $T_f=0.7$ (see SM). As it happens, the vdW approximation for $\mathcal{E}(k;n,T_f)$ locates the critical point (CP) of our system at the state point $(\phi_c,T_c)=(0.13,0.85)$, and not at its actual (simulated) value (0.16,1.22). To scale out these imprecisions, to each simulation state point we assign a theoretical state point $(\phi,T)$ in a linear proportion as the simulation CP relates to the theoretical CP. Within this correspondence, the simulated quench just defined is analogous to the theoretical quench described in  Fig. \ref{Fig1}(a).  The corresponding simulation results for $S(k;t)$ are presented in  Fig. \ref{Fig1}(b).

The comparison between the sequence of simulated snapshots of $S(k;t)$ with the corresponding  theoretical sequence illustrates the qualitative agreement between both description of this non-equilibrium kinetic process. The same comparison also exhibits the quantitative inaccuracies of the approximate theoretical predictions in the early and intermediate stages described by the simulations. Thus, the visual comparison of pairs of snapshots with the same evolution time $t$, indicates a slower evolution of the theoretical $S(k;t)$ compared with the exact evolution represented by the simulations. Also, the position $k_{max} (t)$ of the theoretical small wave-vector peak moves more slowly to the left than in the simulations, so that the predicted growth of  $\xi (t)$ is noticeably slower. These features are also illustrated in the insets of   Fig. \ref{Fig1}.

In spite of this quantitative mismatch between the simulation and the theoretical clocks, we can pair each simulated snapshot with the theoretical snapshot having the same height $S_{max}(t)$ (but, obviously, different evolution time $t$).  As illustrated in the Supplemental Material, this strongly emphasizes the remarkable qualitative similarity between both sequences of snapshots of $S(k;t)$. Let us mention that we observed essentially the same semi-quantitative agreement when comparing our theoretical predictions for the HSAY model with $z=2$ with the BD simulation results for the actual Lennard-Jones system by Lodge and Heyes \cite{heyeslodge}, thus verifying that the main qualitative features of these non-equilibrium structural and morphological processes are independent of the details of the interaction potential.

This prompted us to investigate the extent to which the main qualitative features of this scenario can be recognized in experimental observations where the kinetics of $S(k;t)$ during the arrested spinodal decomposition of colloidal systems with attractive interactions had been recorded. The result is a  general qualitative agreement between the predicted scenario and the experimental observations. This is illustrated already in the inset of Fig. \ref{Fig1}(a), which demonstrates that one can superimpose the experimental data of the evolution of $\xi (t)$ of the full gel-forming process of the globular protein lysozyme in solution reported by Gibaud and Schurtenberger  \cite{gibaud}, on one of the theoretical curves for $\xi (t)$ for our HSAY model, even though the attractive forces in this experimental system are surely much shorter-ranged than our Yukawa attraction with $z=2$, and in spite of the difference in the volume fraction of the respective isochores (theory $\phi=0.08$; experiment $\phi=0.15$).  For the details of this and the following  semi-quantitative comparison, please see the Supplemental Material. In Fig. \ref{Fig4}(a) we expand this comparison, now using similar experimental measurements, this time involving a different globular protein (bovine serum albumin BSA) in solution,  for which the evolution of $\xi (t)$ was reported by Da Vela et al. \cite{davela} for a sequence of quench processes with varying quench depths. Here again the qualitative similarity is remarkable.

\begin{figure}
\includegraphics[scale=.15]{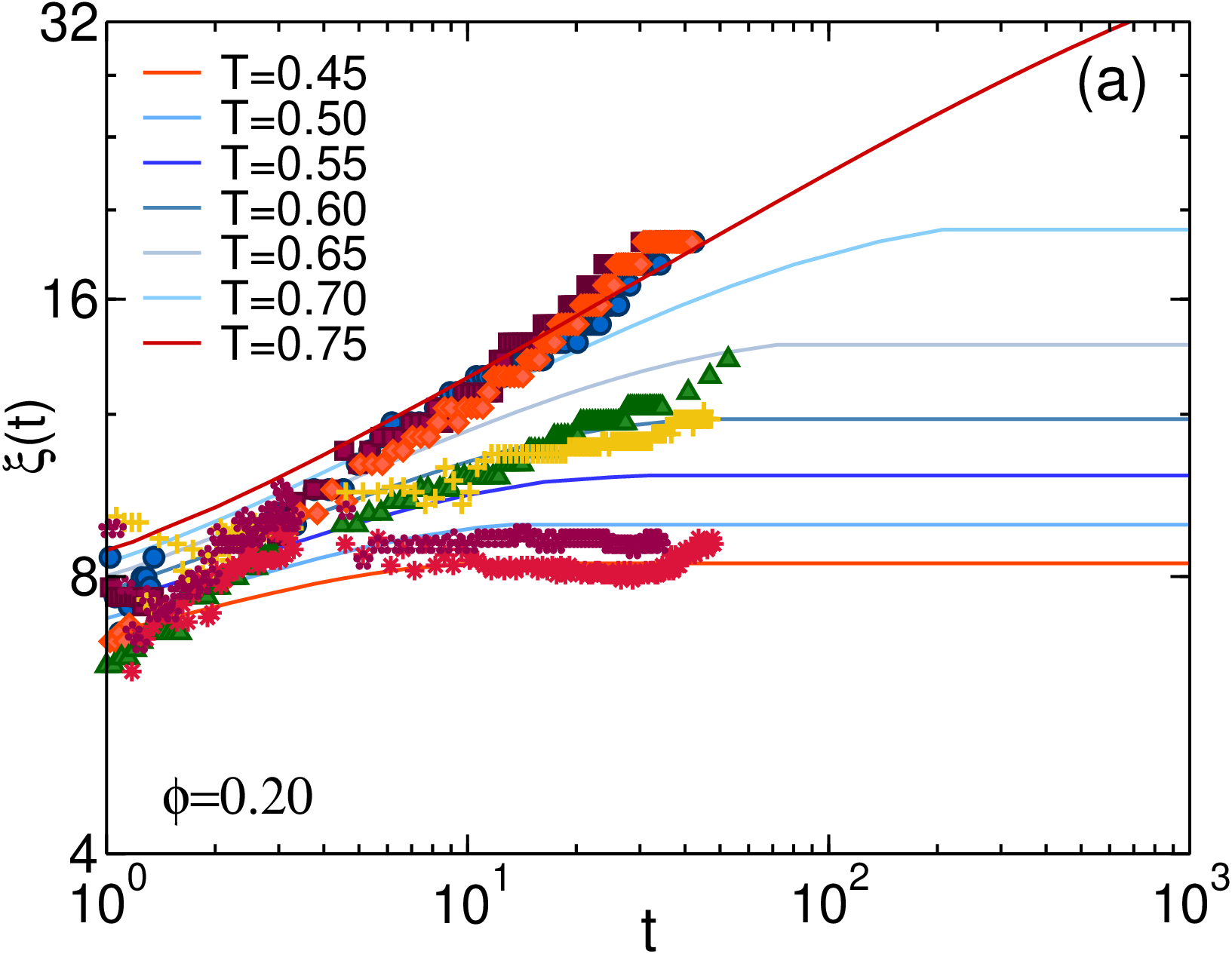}
\includegraphics[scale=.15]{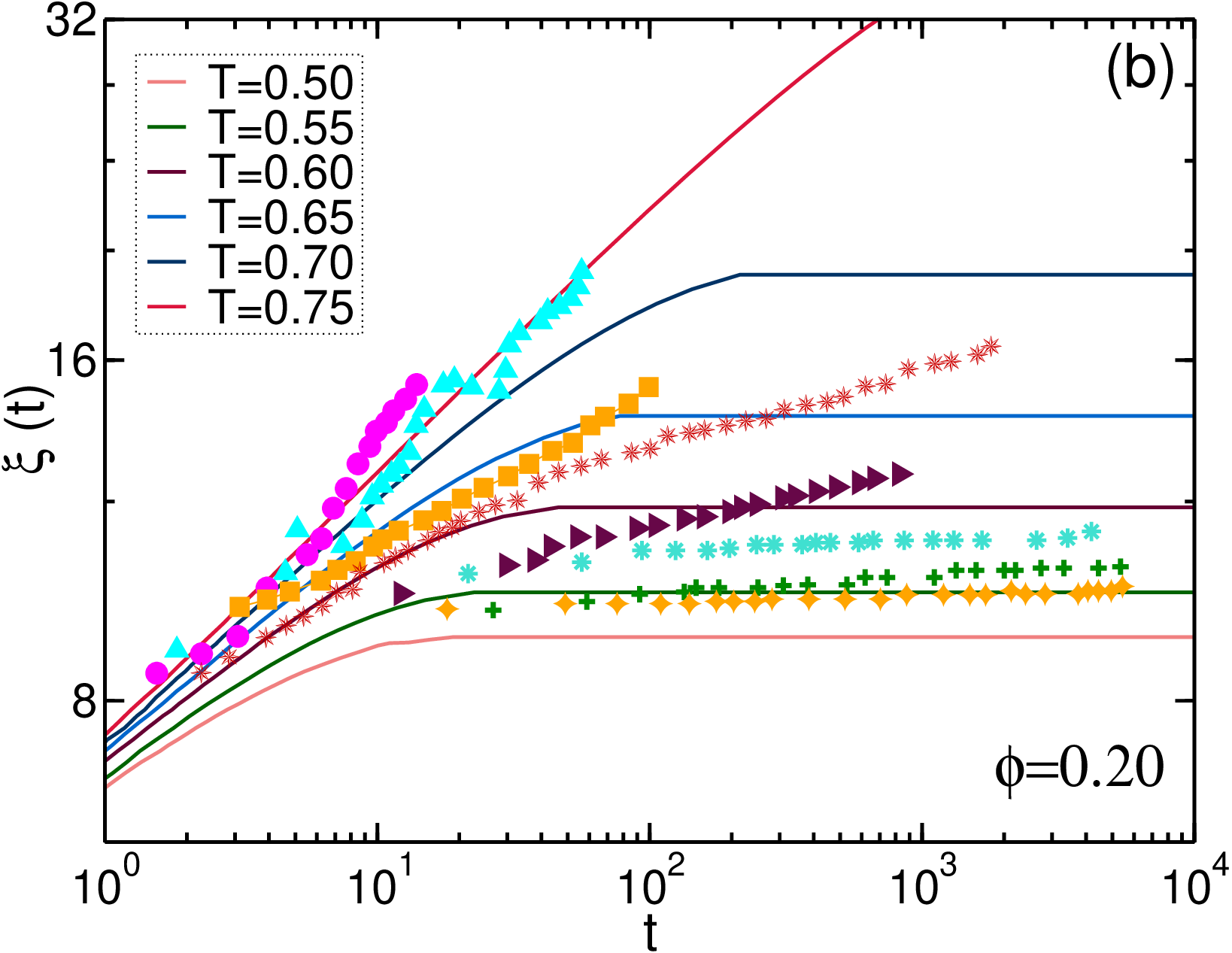}
\caption{(a) Theoretical evolution (lines) of the size $\xi (t)$ for a sequence of quenches for several final temperatures. The symbols are the experimental result measured in the gelling BSA solutions reported in Fig. 7 of Ref. \cite{davela}, scaled by arbitrary factors to illustrate  their  qualitative similarity with the predicted theoretical scenario. (b) Same as (a), but here the experimental results correspond to the measurements in the gelling colloid-polymer mixture with colloid-to-polymer size ratio $\approx 2$ reported in Fig. 5(a) of Ref. \cite{royall}.}
\label{Fig4}
\end{figure}

The range of the attractive interactions between proteins in the previous two examples is much shorter than the HSAY model with $z=2$, but  at least both protein solutions can be regarded as truly  mono-disperse one-component Brownian liquids. The scenario predicted for this model system, however, can also be recognized in systems with still more complex effective attractions, such as in the colloid-polymer mixture studied by Zhang et al. \cite{royall}, in which the colloids attract each other due to polymer-mediated effective depletion forces. This is illustrated in Fig. \ref{Fig4}(b), which corresponds to a colloid-polymer mixture with a ratio of the colloid's radius $\sigma/2$ to the polymer's radius of gyration $R_g$ of $\sigma/2R_g \approx z = 2$, and to a sequence of polymer-concentration quenches at fixed colloid volume fraction $\phi = 0.2$. Once again we observe a remarkable qualitative agreement.

\begin{figure}
\includegraphics[scale=.15]{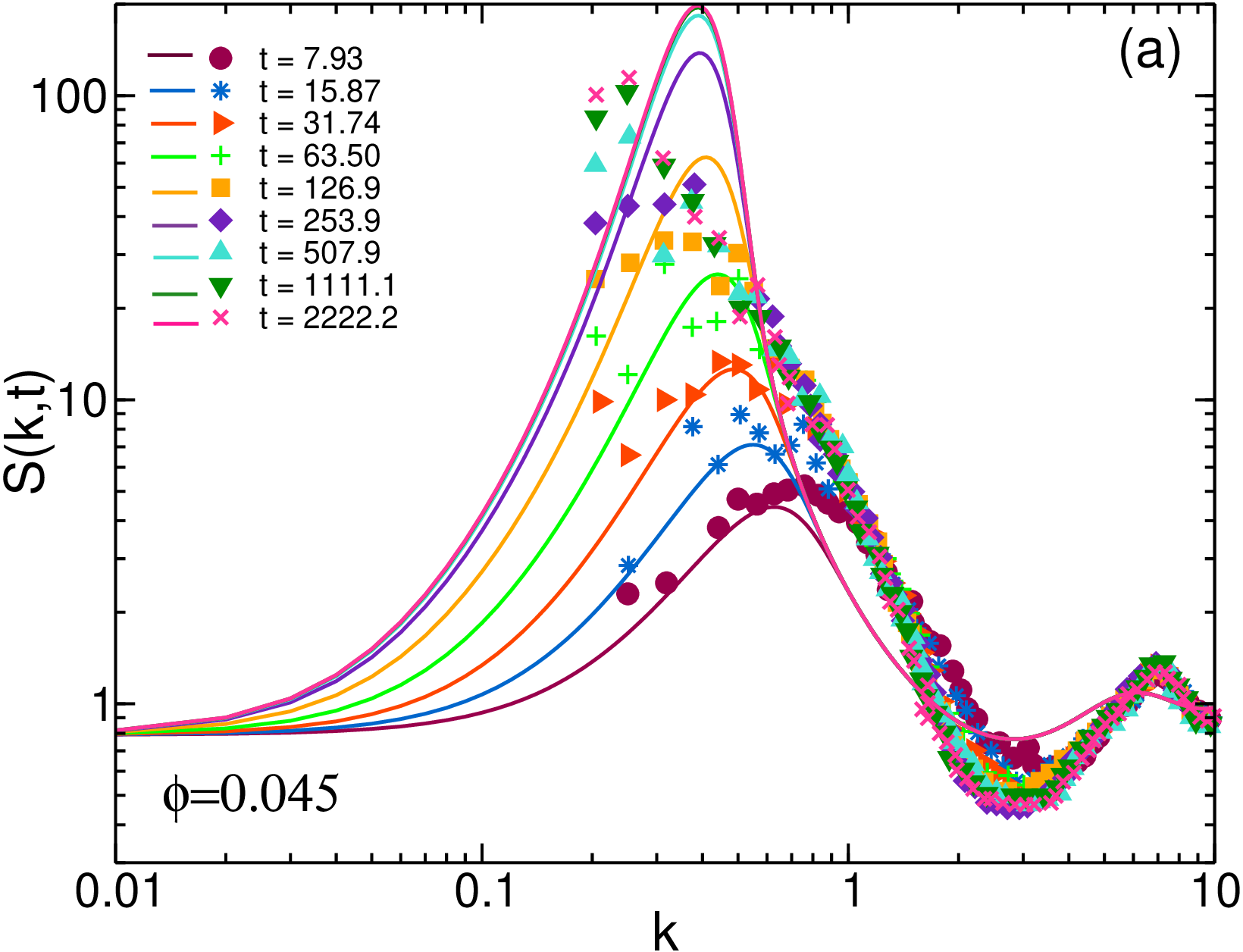}
\includegraphics[scale=.15]{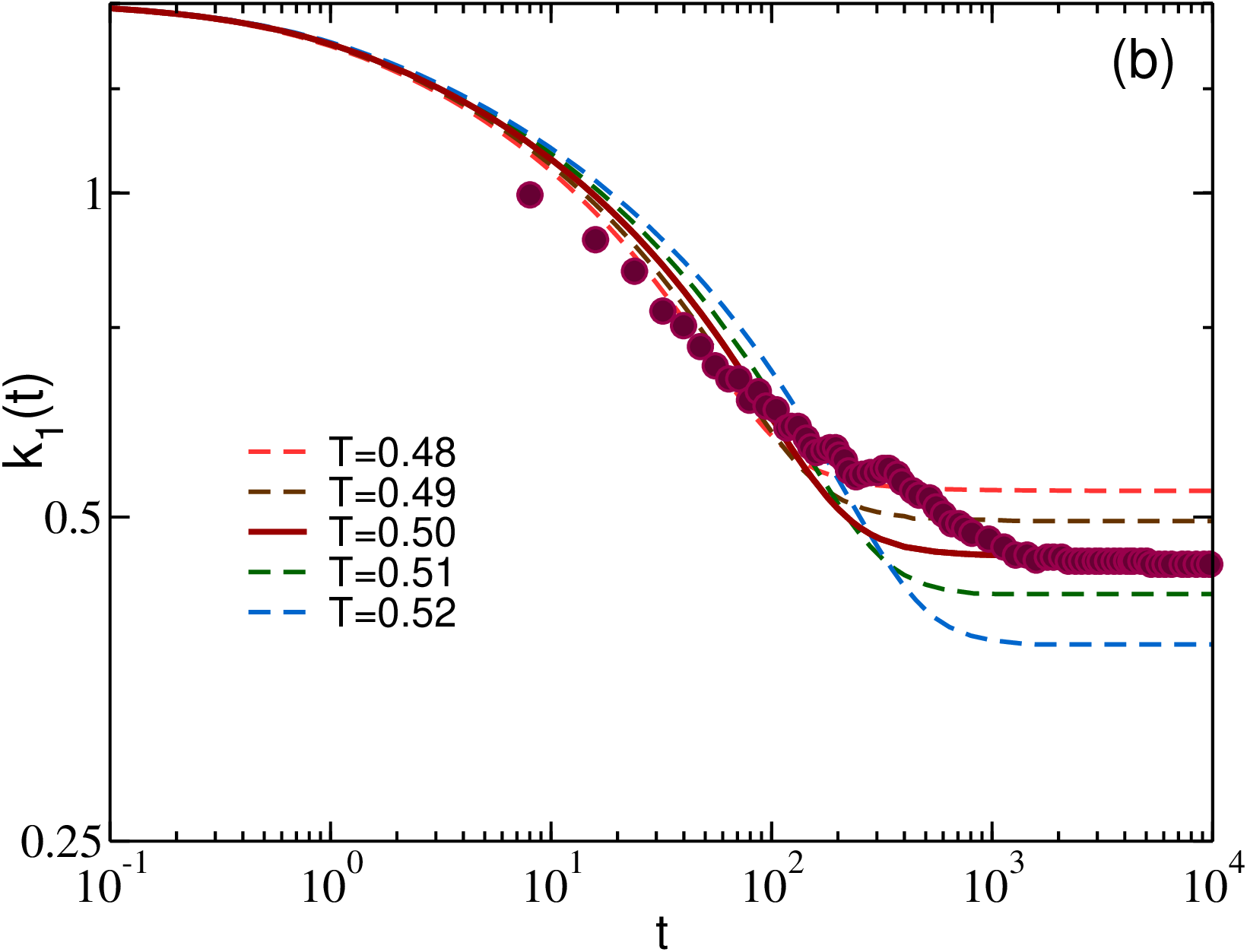}
\caption{(a) Snapshots of the theoretical evolution (lines) and of the experimental measurements (symbols) of $S(k;t)$ in the colloid-polymer mixture  reported in Figs.  4(b) and 4(c) of Ref. \cite{lu}; the initial and final temperature of the theoretical quench are $T_i=5.0$ and $T=0.5$. (b) Evolution of the first moment $k_1 (t)$ of the experimental (symbols) and theoretical (red solid line)  $S(k;t)$ of the quench in (a); the dashed lines correspond to other final temperatures.}
\label{Fig5}
\end{figure}

A similar agreement is observed in Fig. \ref{Fig5} between the predicted and the experimental evolution of the colloid-colloid structure factor $ S(k;t)$ measured by Lu et al. \cite{lu} in a colloid-polymer mixture, this time with a larger colloid to polymer size ratio, $\sigma/2R_g \approx z=17$, and along a more dilute isochore, $\phi=0.045$ (see details in SM). Fig. \ref{Fig5}(a) directly compares snapshots of the structure factor itself and Fig. \ref{Fig5}(b) compares the evolution of its first moment, $k_1(t) \equiv \int_{0}^{k_c} k S(k;t) dk/\int_{0}^{k_c}  S(k;t) dk$, where $k_c$ locates the minimum of $S(k;t)$ after the low-$k$ peak. The wave-vector $k_1(t)$ bears qualitatively the same morphological information as $k_{max}(t)$. Here, too, the qualitative agreement is quite apparent.

This and the other comparisons discussed above illustrate the experimental fact that the arrest of the growth process severely limits the power-law growth regime and makes it strongly dependent on the depth of the quench. This is in contrast with the universality expected from the perspective of theories that do not consider the emergence of dynamic arrest conditions \cite{furukawa}. Instead, what seems to be universal for the class of systems studied here is the non-equilibrium evolution of the full structure factor and of the main morphological parameter ($k_{max}(t)$ or $k_1(t)$) describing the growth and arrest of the spinodal decomposition heterogeneities, the main feature captured by our theory.
Let us emphasize, however, that the solution of the NE-SCGLE equations  renders much more detailed information on the spatio-temporal non-equilibrium  evolution of the structure and dynamics of arresting systems, than those specific features in which we have focused in this communication. Other highly remarkable and counterintuitive kinetic features are concomitant to the previously-discussed growth and arrest of the spinodal heterogeneities. Their adequate discussion, however, deserves more detailed reports that will be communicated separately.

ACKNOWLEDGMENTS: The authors acknowledge Dr. Fajun Zhang and Stefano da Vela for kindly providing the experimental data of Fig. 4(a) and Professor David A. Weitz and Dr. Peter Lu for kindly providing the experimental data of Fig.  5(c) and 5(d). This work was supported  by the Consejo Nacional de Ciencia y Tecnolog\'{\i}a (CONACYT, M\'{e}xico), through grants: 182132, 242364, FC-2015-2-1155 and Laboratorio Nacional de Ingeniería de la Materia Fuera de Equilibrio-279887-2017.


\begin{references}
\bibitem{atomic0} For a colloidal liquid $D_0$ is the short-time self-diffusion coefficient, while for an atomic liquid $D_0$ would be given by Eq. (1) of Ref.  \cite{atomic1}.
\end{references}
\end{document}


\draft
\emph{}
\title{Non-equilibrium Kinetics of the Structural and Morphological Transformation of Liquids into Physical Gels}
\author{Jos\'e Manuel Olais-Govea, Leticia L\'opez-Flores, Mart\'in Ch\'avez-P\'aez, and
Magdaleno Medina-Noyola}

\address{Instituto de F\'{\i}sica {\sl ``Manuel Sandoval Vallarta"},
Universidad Aut\'{o}noma de San Luis Potos\'{\i}, \'{A}lvaro
Obreg\'{o}n 64, 78000 San Luis Potos\'{\i}, SLP, M\'{e}xico}

\date{\today}

\section{Supplemental Material (SM)}

\subsection{The main approximations of the NE-SCGLE theory}

The essence of the NE-SCGLE theory \cite{nescgle1} are the time-evolution equations of: (I) the mean value
$\overline{n}(\textbf{r},t)$,
\begin{equation} \frac{\partial \overline{n}(\textbf{r},t)}{\partial
t} = D_0{\nabla} \cdot b(\textbf{r},t)\overline{n}(\textbf{r},t)
\nabla \beta\mu[{\bf r};\overline{n}(t)], \label{difeqdlp}
\end{equation}
and of: (II) the Fourier transform (FT) $\sigma(k;\textbf{r},t)$ of the covariance $\sigma(\textbf{r},\textbf{r} + \textbf{x};t)$,
\begin{eqnarray}
\begin{split}
\frac{\partial \sigma(k;\textbf{r},t)}{\partial t} = & -2k^2 D_0
\overline{n}(\textbf{r},t) b(\textbf{r},t)
\mathcal{E}(k;\overline{n}(\textbf{r},t)) \sigma(k;\textbf{r},t)
\\ & +2k^2 D_0 \overline{n}(\textbf{r},t)\ b(\textbf{r},t), \label{relsigmadif2p}
\end{split}
\end{eqnarray}
of the fluctuations of the local density $n(\textbf{r},t)$ of particles. In these equations $D_0$ is the particles' short-time self-diffusion coefficient and $b(\textbf{r},t)$ is their local reduced mobility.  The main external input of these equations is the Helmholtz free energy density-functional $\mathcal{F}[n]$, or, more precisely, its first and second  functional derivatives: the chemical potential $\mu [{\bf r};n] \equiv \left[ {\delta \mathcal{F}[n]}/{\delta n({\bf r}')}\right]$ and the thermodynamic function $\mathcal{E}[{\bf r},{\bf r}';n]\equiv \left[ {\delta \beta\mu [{\bf r};n]}/{\delta n({\bf r}')}\right] $. In Eq. (\ref{relsigmadif2p}),  $\mathcal{E}(k;\overline{n}(\textbf{r},t))$ is the Fourier transform (FT) of $\mathcal{E}[{\bf r},\textbf{r} + \textbf{x};n]\equiv \left[ {\delta \beta\mu
[{\bf r};n]}/{\delta n(\textbf{r} + \textbf{x})}\right]_{n=\overline{n}(\textbf{r},t)}$.

In principle, these two equations describe the \emph{isochoric} non-equilibrium morphological and structural evolution of a simple liquid of $N$ particles in a volume $V$ after being \emph{instantaneously} quenched at time $t=0$ to a final temperature $T_f$, in the absence of applied external fields. This description, cast in terms of the one- and two-particles distribution functions  $\overline{n}(\textbf{r},t)$ and $\sigma(k;\textbf{r},t)$, involves  the local mobility $b(\textbf{r},t)$, which is in reality a \emph{functional} of $\overline{n}(\textbf{r},t)$ and $\sigma(k;\textbf{r},t)$, and this introduces strong non-linearities. In fact, even before solving these equations, they reveal a relevant feature of general and universal character: besides the equilibrium stationary solutions $\overline{n}^{eq}(\textbf{r})$ and $\sigma^{eq} (k;\textbf{r})$, defined by the equilibrium conditions $\nabla \beta\mu[{\bf r};\overline{n}^{eq}]=0$ and $\mathcal{E}(k;\overline{n}(\textbf{r},t)) \sigma(k;\textbf{r},t)=1$, Eqs. (\ref{difeqdlp}) and (\ref{relsigmadif2p}) also predict the existence of another set of stationary solutions that satisfy the dynamic arrest condition, $\lim_{t_w\to \infty} b(\textbf{r},t)=0$. This far less-studied second set of solutions describes, however, important non-equilibrium stationary states of matter, corresponding to common and ubiquitous non-equilibrium amorphous solids, such as glasses and gels.

To appreciate the essential physics, the best is to provide explicit examples. To do this at the lowest mathematical and numerical cost, let us write $\overline{n}(\textbf{r},t)$ as the sum of its bulk value $n\equiv N/V$ plus the deviations $\Delta \overline{n}(\textbf{r},t)$ from homogeneity, and in a zeroth-order approximation let us neglect  $\Delta \overline{n}(\textbf{r},t)$. As explained in more detail in Ref. \cite{nescgle6}, this reduces the previous two equations to only one equation for the covariance, now written in terms of the non-equilibrium structure factor $S(k;t)$ as $\sigma(k;\textbf{r},t)=nS(k;t)$. For waiting times $t>0$ after the quench, such an equation reads
\begin{equation}
\frac{\partial S(k;t)}{\partial t} = -2k^2 D_0
b(t)n\mathcal{E}_f(k) \left[S(k;t)
-1/n\mathcal{E}_f(k)\right], \label{relsigmadif2pp}
\end{equation}
where $\mathcal{E}_f(k)\equiv \mathcal{E}(k;n,T_f)$ is the value of $\mathcal{E}(k;\overline{n}(\textbf{r},t))$ at the uniform profile $\overline{n}(\textbf{r},t)= n$ and at the final temperature $T_f$.
Here, too, $b(t)$ is in reality a functional of $S(k;t)$

\subsection{The local mobility $b(t)$ as a functional of $S(k;t)$.}

The detailed functional dependence of the local mobility $b(t)$ on  the non-stationary structure factor $S(k;t)$ is determined by the following NE-SCGLE equations, which must be self-consistently solved together with Eq. (\ref{relsigmadif2pp}). As explained in Ref. \cite{nescgle2}, this set of equations start by writing $b(t)$ as
\begin{equation}
b(t)= [1+\int_0^{\infty} d\tau\Delta{\zeta}^*(\tau; t)]^{-1},
\label{bdt}
\end{equation}
with the $t$-evolving, $\tau$-dependent friction function $\Delta{\zeta}^*(\tau; t)$ given approximately by
\begin{equation}
\begin{split}
  \Delta \zeta^* (\tau; t)= \frac{D_0}{24 \pi^{3}n} \int d^{3} k\ k^2 \left[\frac{ S(k;t)-1}{S(k; t)}\right]^2  \\ \times F(k,\tau; t)F_S(k,\tau; t),
\end{split}
\label{dzdtquench}
\end{equation}
in terms of $S(k; t)$ and of the collective and self non-equilibrium intermediate scattering functions $F(k,\tau; t)$ and $F_S(k,z; t)$, whose memory-function equations are written approximately, in terms of
the Laplace transforms $F(k,z; t)$ and $F_S(k,\tau; t)$, as

\begin{gather}
\label{fluctquench}
 F(k,z; t) = \frac{S(k; t)}{z+\frac{k^2D_0 S^{-1}(k;
t)}{1+\lambda (k)\ \Delta \zeta^*(z; t)}},
\end{gather}
and
\begin{gather}\label{fluctsquench}
 F_S(k,z; t) = \frac{1}{z+\frac{k^2D_0 }{1+\lambda (k)\ \Delta
\zeta^*(z; t)}},
\end{gather}
with $\lambda (k)$ being a phenomenological interpolating function \cite{todos2},
\begin{equation}
\lambda (k)=1/[1+( k/k_{c}) ^{2}], \label{lambdadk}
\end{equation}
in which $k_c$ is an empirically determined parameter (here we use $k_c$ as $k_c= 1.305 \times 2\pi/\sigma$, as in previous works).

Eqs. (\ref{relsigmadif2pp})-(\ref{lambdadk}) summarize the NE-SCGLE theory employed so far to describe the irreversible processes occurring in a solidifying glass- or gel-forming liquid. A systematic presentation of the predictions of this theory and of their correspondence with the widely observed experimental signatures of the glass transition, started in Refs. \cite{nescgle3} and \cite{nescgle6} with the description of the transformation of equilibrium hard-sphere (and soft-sphere) liquids into ``repulsive'' glasses. That investigation was extended in Ref. \cite{nescgle5} to Lennard-Jones--like simple liquids (pairwise interactions composed of a strong repulsion plus an attractive tail), which revealed a much richer scenario, summarized by the \emph{non-equilibrium} phase diagram in Fig. 1 of our Letter. The scenario laid down in Ref. \cite{nescgle5} was inferred solely on the basis of the predicted long-time asymptotic stationary solutions of the NE-SCGLE equations.

\subsection{Model system and approximate thermodynamic input.}

For a monocomponent simple liquid with pairwise interaction $u(r)=u_{HS}(r) + u_{A}(r)$ (hard spheres of diameter $\sigma$, plus a weaker attractive interaction $u_{A}(r)$), we shall adopt the van der Waals (vdW) approximation for  the Helmholtz free energy functional, $\mathcal{F}[n]=\mathcal{F}_{HS}[n]+ \frac{1}{2}\int d\mathbf{r} d\mathbf{r}' n(\textbf{r})n(\textbf{r}') u_{A}(|\textbf{r}-\textbf{r}'|)\theta (|\textbf{r}-\textbf{r}'|-\sigma)$, where $\mathcal{F}_{HS}[n]$ is the exact free energy functional of the reference HS system and $\theta(x)$ is Heaviside's step function. This leads to the  approximate chemical potential, $\mu [{\bf r};n] = \mu_{HS} [{\bf r};n] + \int d\mathbf{r}'  u_{A}(|\textbf{r}-\textbf{r}'|)\theta (|\textbf{r}-\textbf{r}'|-\sigma)n(\textbf{r}')$, and thermodynamic functional
$\mathcal{E}[{\bf r},{\bf r}';n]=\mathcal{E}_{HS}[{\bf r},{\bf r}';n]+ \beta u_{A}(|\textbf{r}-\textbf{r}'|)\theta (|\textbf{r}-\textbf{r}'|-\sigma)$, which in Fourier space reads $\mathcal{E}(k;n)=\mathcal{E}_{HS}(k;n)+ \beta u_{A}(k)$. This last equation was referred to  in Ref. \cite{nescgle5} as Sharma-Sharma approximation \cite{sharmasharma}.

The hard-sphere function $\mathcal{E}_{HS}(k;n)$ can be determined using the Ornstein-Zernike (OZ) equilibrium condition for the static structure factor, $S^{eq}(k;n)=1/n\mathcal{E}_{HS}(k;n)$. This OZ equation, complemented with the Percus-Yevick approximation \cite{percusyevick} with Verlet-Weis correction  \cite{verletweis}, provides an analytic expression for $\mathcal{E}_{HS}(k;n)$. The van der Waals approximate free energy, complemented by these (virtually exact) hard-sphere  properties, were employed to solve Eqs. (\ref{relsigmadif2pp})-(\ref{fluctsquench}) for the HSAY model. The spinodal curve in Fig. 1  of the Letter, was obtained from the condition $\mathcal{E}(k=0;\phi,T_s)=0$.

\subsection{Simulations.}
Brownian dynamics simulations were performed using the algorithm
developed by Ermak and Mcammon \cite{tildesley} without hydrodynamics interactions,
using a cubic simulation box, periodic boundary conditions, and a number
of particles $N=4300$. To mimic the hard-core part of the attractive
Yukawa potential, a short-range repulsive part was considered. In specific,
the simulations employed the soft-sphere plus attractive Yukawa form
\begin{equation}
u(r)=
\epsilon\Big[\Big(\frac{\sigma}{r} \Big)^{2\nu}-2\Big(\frac{\sigma}{r} \Big)^{\nu}+1\Big]
\theta(\sigma-r)
-\epsilon \frac{\exp[-z(r/\sigma -1)]}{(r/\sigma )}
\label{yukawa}
\end{equation}

\noindent where $\theta(x)$ is the Heaviside's step function and $\nu=20$.
A cut-off radius $r_c=3.5\sigma$ was employed for the attractive part.
All the simulations were started from a random configuration, equilibrated
for $5\times 10^5$ time steps of length $\Delta t=5.0\times 10^{-5}t_B$ $(t_B \equiv \sigma^{2}/D_{0})$ and using
an initial temperature $T_i=2.0$. A quench was then applied to the
desired temperature  $T_f$, continuing on  the simulations using
a reduced time step $\Delta t=10^{-5}t_B$ and collecting data every 200
time steps. Following Ref. \cite{heyeslodge}, structural functions were calculated
as the averages in non-overlapping time-windows of length $0.2t_B$.
The reported results correspond to the average over ten independent
simulations for every system.

\begin{figure}
\begin{center}
\includegraphics[scale=.25]{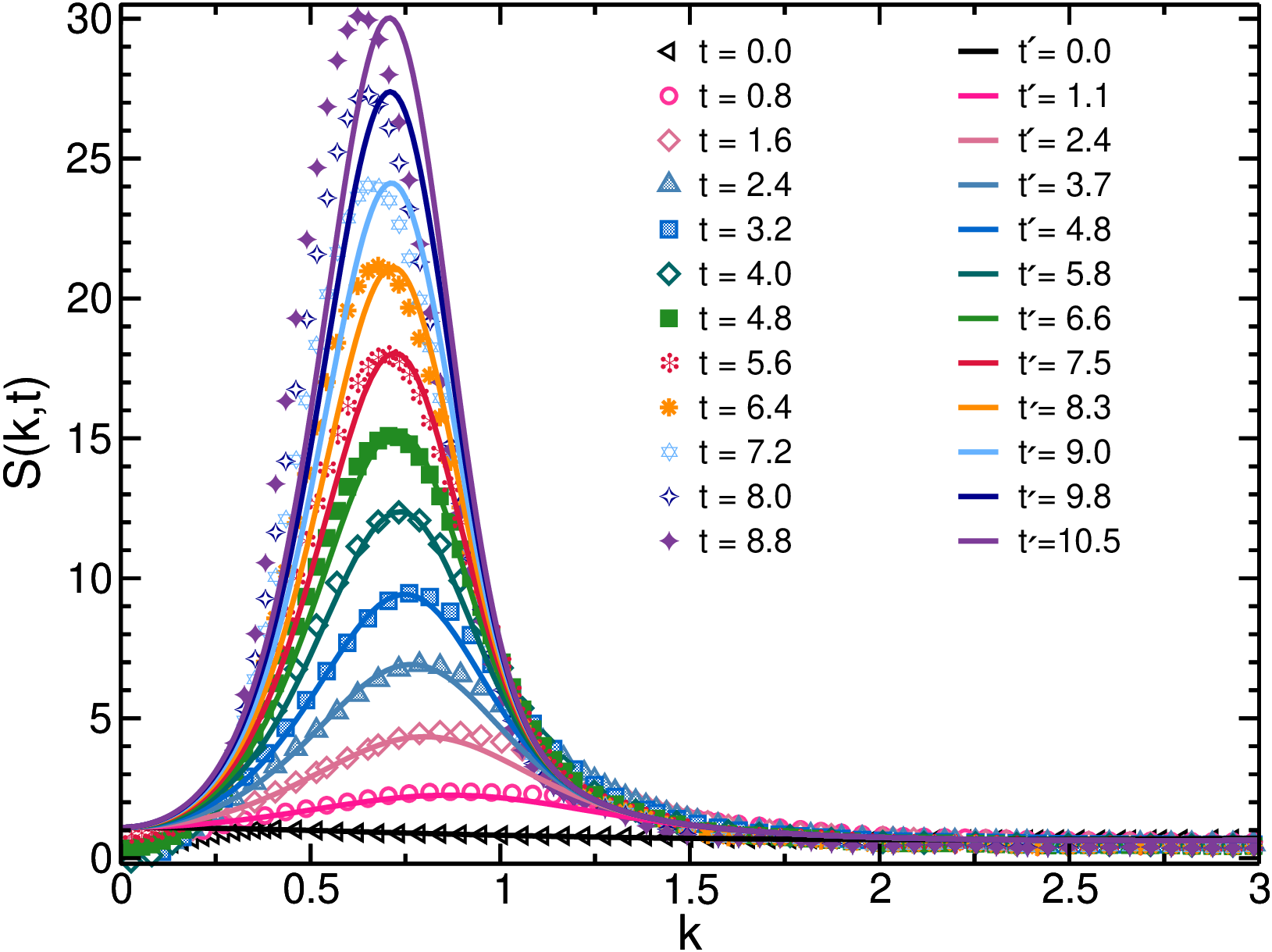}
\caption{Theoretical snapshots (solid lines) of the small-$k$ peak of  $S(k;t)$ (for the same conditions indicated in Fig. 2 of our Letter), compared with the snapshots of a Brownian-dynamics simulated quench (symbols) and with waiting times adjusted to match the height $S(k_{max};t)$ of this peak.} \label{Fig2}
\end{center}
\end{figure}

In Fig. 2 of our Letter, the sequence of simulated snapshots of $S(k;t)$ was compared with the corresponding  theoretical sequence, with identical evolution times $t$, thus exhibiting the quantitative inaccuracies of the approximate theory, most notably a quantitative mismatch between the simulation and the theoretical clocks. In spite of this difference, however, the structural pathways predicted by theory and registered by simulations are remarkably similar. This is illustrated here in Fig. SM1, which compare the same sequence of simulated snapshots of $S(k;t)$ in Fig. 2 of our Letter, but now paired with the theoretical snapshots having the same height $S_{max}(t)$ (but, obviously, different evolution time).

\subsection{Arrested spinodal decomposition in protein solutions.}

Here we provide the quantitative details of the comparisons in Figs. 3,  4(a)-(b) and 5(a)-(b) of the Letter, between our theoretical predictions and the reported experimental measurements of the growth and arrest of the spinodal heterogeneities of gelling protein solutions.

\subsubsection{Gelling lysozyme solutions.}

In Fig. 2(d) of Ref. \cite{gibaud}, Gibaud and Schurtenberger report the experimental measurements of the growth and arrest of the representative size $\xi(t)$ (in micrometers) of the heterogeneities, as a function of time (in seconds), of gelling lysozyme proteins of diameter $\sigma=3.4nm$ in aqueous  solution at fixed bulk concentration corresponding to a volume fraction of approximately $\phi=0.15$. To compare our predicted theoretical scenario with these data, let us normalize their experimental $\xi(t)$ and $t$ obtained by microscopy, by the experimental value of our theoretical units of length and time, $\sigma$ and $\tau_{B}\equiv\sigma^{2}/D_{0}$. Using Stokes-Einstein's relation and the viscosity of water at room temperature we obtain $D_{0}=122\mu m^{2}/s$ and $\tau_{B}= \sigma^{2}/D_{0}=1\times 10^{-7}s$.

\begin{figure}
\includegraphics[scale=.25]{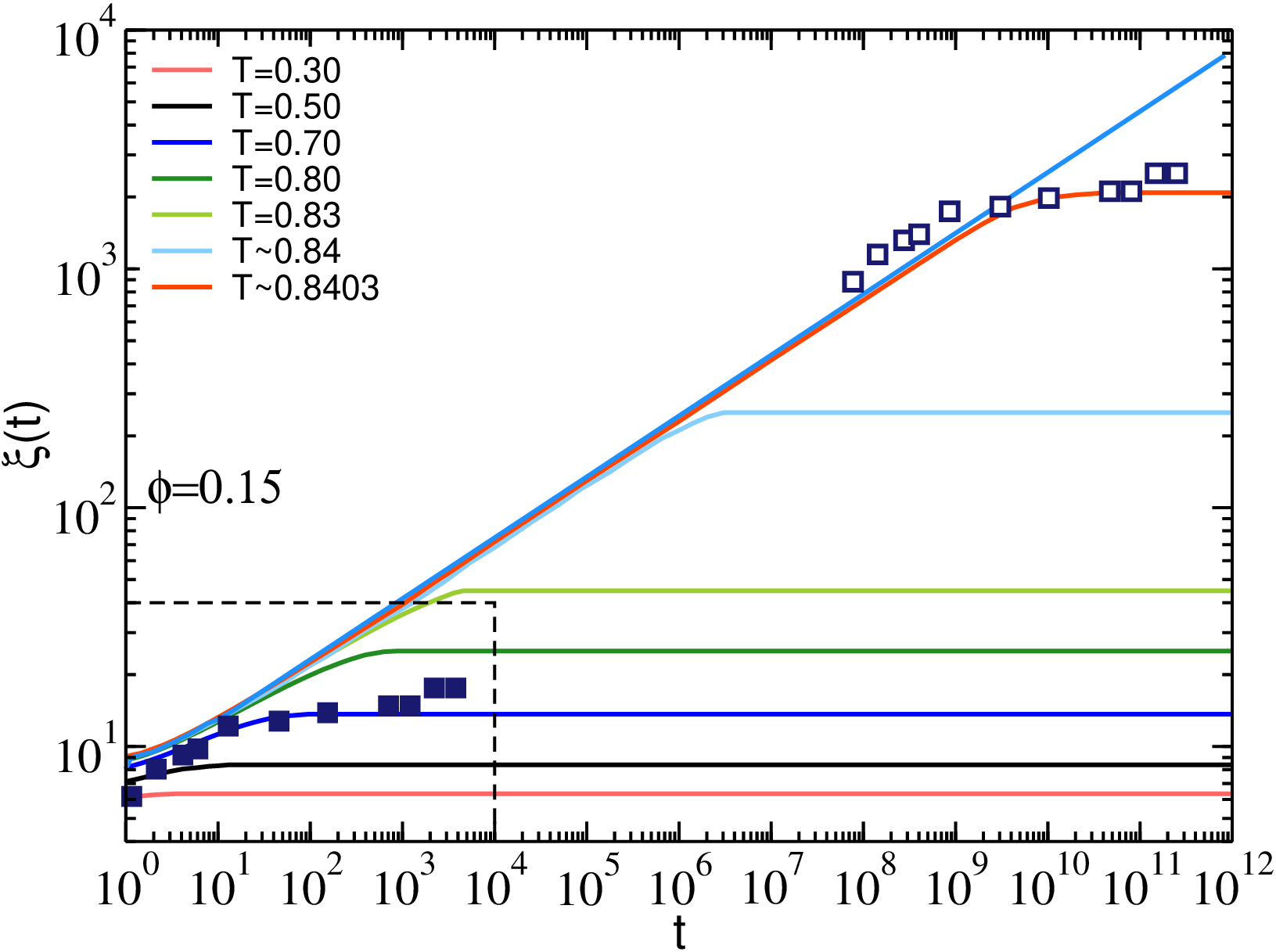}
\caption{Theoretical evolution (solid lines) of the size $\xi (t)$ of the heterogeneities of the HSAY model for a sequence of quenches for several final temperatures. The empty square symbols are the experimental results measured in a lysozyme protein system reported in Fig. 2(d) of Ref.\cite{gibaud}. The solid symbols represent the scaled data.}
\label{Suplementary1}
\end{figure}

Plotted in this manner, the experimental data of Fig. 2(d) of Ref. \cite{gibaud} appear as the empty symbols in Fig. SM \ref{Suplementary1}. In this figure the solid lines are the theoretical predictions for the evolution of $\xi(t)$ in the HSAY model with $z=2$ initially at the (theoretical) temperature $T_0=1.5$ and quenched at fixed volume fraction ($\phi=0.15$) to various values of the final temperature $T$. The region limited by the dotted lines is the time-window employed in the inset of Fig. 2(b). As we can see, in reality the experimental data fall well outside such time-window. However, as our present comparison demonstrates, one can find a theoretical curve (red line) corresponding to a value of  $T$ closer to the spinodal curve, whose predicted arrest also occurs far outside this window and approximately superimposes on the experimental data. The Fig. SM 2 also re-plots, now as solid symbols, the same experimental data with $\xi(t)$ and $t$ arbitrarily reduced by empirical factors (7 $\times 10^{-3}$ and 1.5 $\times 10^{-8}$, respectively). We do this only to illustrate, as we do in the inset of Fig. 2(b) of the Letter, their qualitative resemblance with the predicted scenario of the demonstrative quench of the HSAY model  discussed in that figure (notwithstanding the fact that the theoretical curves there actually refer to the isochore $\phi=0.08$).

Of course, comparing theoretical results for the HSAY model with experimental data for the referred protein solution can only have the purpose of comparing the essential qualitative features of the striking arrest of the process of spinodal decomposition. To establish a more proper quantitative comparison, the theoretical calculations should be fine-tuned to actually correspond to the detailed experimental conditions (much smaller range of the attractive potential ($z>> 2$), finite cooling rates, etc.), which is for the moment out of the scope of this short communication.

\subsubsection{Gelling bovine serum albumin solutions.}

To complement the previous comparison, let us now refer to Fig. 7 of reference \cite{davela}, where Da Vela et al. report quite similar experimental measurements of $\xi(t)$ for \emph{a sequence} of quenches involving  another protein, namely, bovine serum albumin, quenched along its critical isochore (protein concentration $175mg/mL$ BSA with 44 mM YCL$_3$). This protein has diameter $\sigma=6 nm$, and in the reported solution presents a lower consolute temperature. Thus, in contrast with the previous example, a deeper quench involves a higher experimental temperature $T_e$. Nevertheless, to compare these measurements with our predicted scenario we followed essentially the same procedure described above, and the result is summarized in Fig. 3(a) of the Letter.

\begin{figure}
\includegraphics[scale=.25]{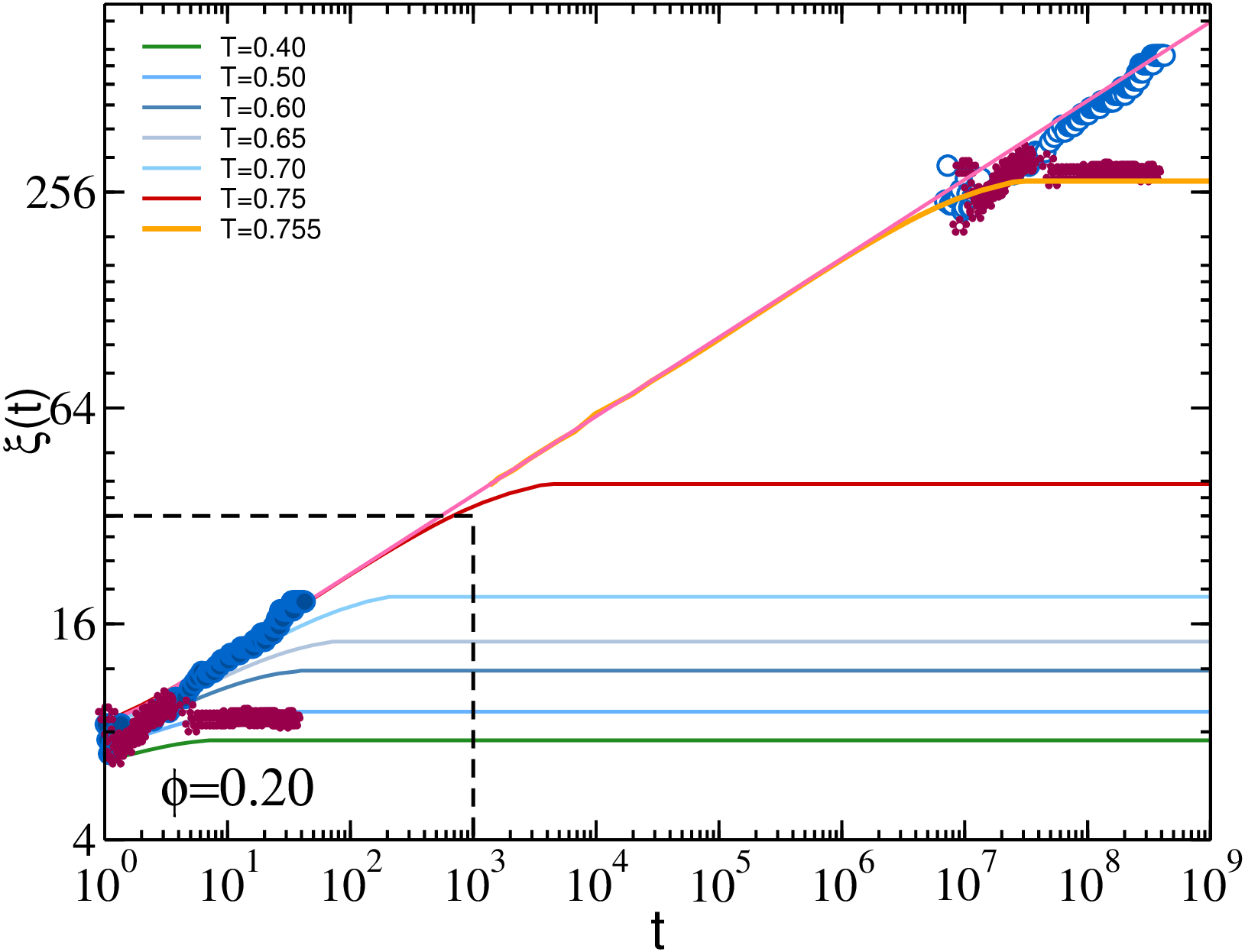}
\caption{ Theoretical evolution (solid lines) of the size $\xi (t)$ of the heterogeneities of the HSAY model for a sequence of quenches for several final temperatures. The symbols are the experimental results measured in the BSA solutions reported in Fig. 7 of Ref. \cite{davela}, for two of the systems (squares for $T=30^{o}C$ and circles for $T=57.5^{o}C$). The solid symbols represent the scaled data.}
\label{Suplementary2}
\end{figure}

To explain this procedure in more detail, let us mention that in this case we estimated $D_{0}= 60 \mu m^{2}/s$ and $\tau_{B}=1\times 10^{-6}s$. In Fig. SM \ref{Suplementary2}, we thus plot the normalized experimental data (empty symbols) corresponding to the shallowest and to the deepest quenches reported  in Fig. 7 of reference \cite{davela}  (labeled there by the experimental temperatures $T_e=30^{o}C$ and $57.5^oC$, respectively). The solid lines are theoretical predictions for the evolution of $\xi(t)$ in the HSAY model with $z=2$ initially at the (theoretical) temperature $T_0=1.5$ for various values of the final temperature $T$. Since the experiments were conducted at the volume fraction $\phi = 0.20$, the theoretical quenches were performed at the same volume fraction.

Once again, the experimental data fall well outside the time-window employed in the inset of Fig. 2b, but again, one can find theoretical curves corresponding to values of  $T$ closer to the spinodal curve, whose predicted arrest approximately superimposes on these experimental data. This Fig. SM3 also re-plots (solid symbols) the same experimental data with $\xi(t)$ and $t$ arbitrarily reduced by empirical factors (3.1 $\times 10^{-2}$ and 1 $\times 10^{-7}$, respectively). This is exactly what we have done to generate Fig. 3(a), which also includes the experimental results for the other quenches reported in Fig. 7 of reference \cite{davela}.

\subsection{Arrested spinodal decomposition in colloid-polymer mixtures.}

Let us now discuss the details of the comparisons in Figs. 3(b), 4(a)-(b) of the Letter, which refer to the experimental measurements of the evolution of $\xi (t)$ in two colloid-polymer mixtures that differ in the range of the depletion attraction between colloids.

\subsubsection{Moderate polymer/colloid size ratio: longer-ranged depletions.}

In Fig. 3(b) of the Letter we quote the experimental measurements by Zhang et al., reported in Fig. 5(a) of Ref. \cite{royall}, of the growth and arrest of $\xi(t)$ in a colloid-polymer mixture where  the mean diameter of the colloids is 544 nm and the polymer radius of gyration is 126 nm, so that the range $\delta$ of the depletion forces between colloids induced by the polymer  (in units of the colloid's diameter) is $\delta\approx 0.45$. This means that, regarding the range of the attractive interactions, this system is represented more closely (than the previous protein solutions) by the HSAY model with $z = 2$, discussed and simulated in Fig. 2 of our Letter.

Ref. \cite{royall} determines that for these experimental systems $\tau_{B} \approx 0.7s$ and reports the data already in units  of  $\sigma$ and $\tau_{B}$ for a system with fixed colloid volume fraction $\phi=0.20$ and for several values of the polymer concentration $C_p$, whose inverse plays the role of the theoretical temperature $T$. In Fig. 3(b) of the Letter we plot these experimental data along with our theoretical predictions for a sequence of quenches along the isochore $\phi=0.20$. Here again, to put both theory and experiments in the same time window, so as to appreciate the qualitative agreement, the experimental data of $\xi(t)$ and $t$ were arbitrarily multiplied by empirical factors (6 and 9, respectively),  much more moderate than in the previous case involving  protein solutions.

\subsubsection{Small polymer/colloid size ratio: short-ranged depletions.}

Figs. 4(a) and (b) of our Letter reproduce, respectively, the experimental data reported by Lu et al. in Figs. 4(b) and (c) of Ref. \cite{lu}, corresponding to a colloid-polymer mixture with colloid radius 560 nm and colloid-to-polymer size ratio $z= \sigma/2R_g \approx17$. In these experiments, the transition to arrested or gelled states is investigated as a function of the polymer concentration, which plays the role of an effective temperature. These experiments monitor the non-equilibrium evolution of the colloid-colloid static structure factor $S(k;t)$ and of its first moment $k_1(t) \equiv \int_{0}^{k_c} k S(k;t) dk/\int_{0}^{k_c}  S(k;t) dk$ (where $k_c$ locates the minimum of $S(k;t)$ following its low-$k$ peak), after a quench from a polymer concentration $c_p= 0$ to $c_p= 53.31$ mg/ml, keeping the colloid volume fraction fixed at $\phi=0.045$.  The viscosity of the solvent is reported to be $\eta=1.96\times 10^{-3}$Pa $\cdot$ s at the temperature $T=25^{o}$C. Using the Stokes-Einstein relation we obtain the short time self diffusion coefficient as $D_{0}=0.19\mu m^{2}/s$, yielding a characteristic time $\tau_{B}= \sigma^{2}/D_{0} = 6.3s$, which is the time unit employed in Figs. 4(a) and (b) of our Letter. The measured wave-vector $k_1(t)$ decreases with time until reaching the stationary arrested value $k_{1}^{a}\approx 0.45 $.

To establish  the connection between theory and experiments, we first solved the  NE-SCGLE Eqs. (\ref{relsigmadif2pp})-(\ref{lambdadk}) to theoretically calculate the first moment $k_{1}(t)$ for a sequence of quenches along the isochore $\phi=0.045$ of the HSAY model ($z=2$), starting at the same (high) initial temperature and with varying (lower) final temperature $T$. We then determined the $T$-dependent limiting value  $k_{1}^{a} =\lim_{t \to\infty} k_{1}(t)$, and chose as the effective temperature of the experimental quench, the value of $T$ for which the condition $k_{1}^{a}(T) \approx 0.45 $ was satisfied, yielding $T \approx 0.5$. In Fig. 4(a) of the Letter we compare the solution of Eqs. (\ref{relsigmadif2pp})-(\ref{lambdadk}) for $S(k;t)$ after this particular quench, with the corresponding experimental data. Fig. 4(b) of the Letter presents a similar comparison between the theoretical (thick solid line) and experimental (full circles) time-dependent first moment $k_{1}(t)$.


The direct comparison between theory and experiment presented in Fig. 4 of the Letter illustrates that, in spite of appreciable quantitative discrepancies, most notably the mismatch between the theoretical and experimental clocks, there is a remarkable qualitative similarity. 